\documentclass[twocolumn,prl,superscriptaddress,showpacs]{revtex4}
\usepackage{graphicx}
\usepackage{times}
\usepackage{amsmath,amssymb}

\newcommand{\bra}[1]{\left\langle{#1}\right|}
\newcommand{\ket}[1]{\left|{#1}\right\rangle}
\newcommand{\proj}[1]{\mathcal{P}_{#1}}

\begin{document}

\title{Collective states of interacting Fibonacci anyons}

\author{Simon Trebst}
\affiliation{Microsoft Research, Station Q, University of California, 
Santa Barbara, California 93106}
\author{Eddy Ardonne}
\affiliation{California Institute of Technology, Pasadena, California 91125} 
\affiliation{Nordita, Roslagstullsbacken 23, SE-106 91 Stockholm, Sweden}
\author{Adrian Feiguin}
\affiliation{Microsoft Research, Station Q, University of California, 
Santa Barbara, California 93106}
\author{David A. Huse}
\affiliation{Department of Physics, Princeton University,
Princeton, New Jersey 08544}
\author{Andreas W. W. Ludwig}
\affiliation{Physics Department, University of California, Santa
Barbara, California 93106}
\author{Matthias Troyer}
\affiliation{Theoretische Physik, Eidgen\"ossische Technische
Hochschule Z\"urich, 8093 Z\"urich, Switzerland}

\date{\today}

\begin{abstract}
We show that chains of interacting Fibonacci anyons can support a
wide variety of collective ground states ranging from extended
critical, gapless phases to gapped phases with ground-state
degeneracy and quasiparticle excitations. In particular, we
generalize the Majumdar-Ghosh Hamiltonian to anyonic degrees of
freedom by extending recently studied pairwise anyonic
interactions to three-anyon exchanges. The energetic competition
between two- and three-anyon interactions leads to a rich phase
diagram that harbors multiple critical and gapped phases.
For the critical phases and their higher symmetry endpoints
we numerically establish descriptions in terms of two-dimensional
conformal field theories. A topological symmetry protects the
critical phases and determines the nature of gapped phases.
\end{abstract}



\maketitle


Two-dimensional topological quantum liquids such as the fractional
quantum Hall (FQH) states harbor exotic quasiparticle excitations
which due to their unusual exchange statistics are referred to as
anyons.  Interchanging two anyons may result in not only a
fractional exchange phase, but may also give rise to a unitary
rotation of the original wave function in a degenerate
ground-state manifold. This latter case of non-Abelian statistics
is proposed to be exploited in the context of
topological quantum computation 
\cite{Kitaev03,DasSarma07}.  Intense experimental efforts
\cite{Marcus07} are currently under way to demonstrate the
non-Abelian character of quasiparticle excitations in certain FQH
states, as proposed theoretically \cite{MooreRead,ReadRezayi}.

Given a set of several non-Abelian anyons we can ask what kind of
collective states are formed if these anyons are {\em
interacting} with each other.
A first step in this direction has recently been
taken by studying chains of ``Fibonacci anyons" with
nearest-neighbor interactions \cite{GoldenChain}. Fibonacci anyons
represent the non-Abelian part of the quasiparticle statistics in
the $k=3$ $Z_k$-parafermion `Read-Rezayi' state \cite{MooreRead}, 
an effective theory for FQH liquids at filling fraction $\nu=12/5$
\cite{Xia04}.  A single Fibonacci anyon carries a topological
charge $\tau$.  Two such anyons may combine (``fuse'') so the pair
has charge $\tau$ or has no charge, which is denoted $1$. This is
analogous to two $SU(2)$ spin-1/2's combining to 
either spin-1 or spin-zero total spin.
A two-anyon interaction assigns different energy to the two 
possible charges of the pair, just as a Heisenberg exchange 
interaction does for the two possible total values of spin
of a pair of $SU(2)$ spin-1/2's.
For a chain of Fibonacci anyons with a uniform pairwise
nearest-neighbor interaction of either sign it has been explicitly
shown \cite{GoldenChain} that the Hamiltonian has a topological 
symmetry, which was predicted to stabilize one of the gapless 
phases.
In this manuscript, we give a broader perspective on possible 
collective phases of interacting Fibonacci anyons and 
phase transitions between them.

\begin{figure}[b]
  \includegraphics[width=\columnwidth]{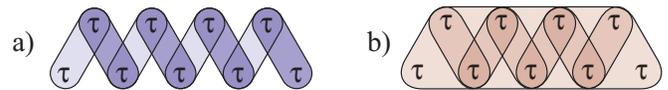}
  \caption{
    (color online) Illustration of
    a) the alternating chain and
    b) the generalized Majumdar-Ghosh chain
         with a three-anyon interaction term.
    Shaded enclosures indicate the fusion products that are energetically
    biased by the Hamiltonian.
   }
  \label{Fig:Models}
\end{figure}

The specific model we will focus on has, in addition to the
two-anyon term, an additional three-anyon interaction
both of which may arise from tunneling \cite{MGfootnote}. 
We find a rich ground state phase diagram that harbors multiple 
critical, gapless and gapped phases. We also mention in passing 
other one-dimensional models that we have investigated and will 
report on in a forthcoming longer paper \cite{long}.  
The topological symmetry, introduced in Ref. \onlinecite{GoldenChain},
that measures the topological flux through a ring of Fibonacci
anyons plays an essential role in determining
the nature of the observed phases and phase transitions.
In particular, we find that this topological symmetry {\em protects}
all the critical phases against spatially uniform local
perturbations. These extended critical phases can be described in
terms of 2D conformal field theories (CFT) with central charges
$c=7/10$ and $c=4/5$ and can be respectively mapped exactly onto
the tricritical Ising and 3-state Potts critical points of the
generalized hard hexagon model \cite{GoldenChain, Huse}. At the
phase transitions out of the tricritical Ising phase into adjacent gapped
phases the system exhibits even higher symmetries which we
identify as {\it tetra}critical Ising and 3-state Potts critical points. 
Remarkably, this demonstrates that these 2D
classical models share an identical non-local symmetry which is
the classical analog of the topological symmetry in the 1D quantum
chains. At the transition into the gapped phases this topological
symmetry is spontaneously broken which results in a non-trivial
ground-state degeneracy in the gapped phases.


\paragraph{Two-anyon interactions.--}

For a uniform chain of Fibonacci anyons the Hamiltonian
introduced in Ref.~\onlinecite{GoldenChain} energetically 
favors one or the other of the possible fusion products of two
neighboring $\tau$-particles which, by the fusion rule $\tau
\times \tau = 1 + \tau$, can be either a $1$ or a $\tau$. 
The energy of the former is lower for a coupling
that is termed `antiferromagnetic' (AFM),
in analogy to the familiar $SU(2)$ spins-chains,
while that of the  latter is lower with a coupling termed `ferromagnetic' (FM).
The underlying Hilbert space is spanned by an orthonormal basis of states, 
each state corresponding to one possible labeling of the chain \cite{GoldenChain}
of repeated fusions with $\tau$. Each site along this chain of fusions
has either a $1$ or a $\tau$, with a constraint forbidding
two adjacent $1$'s. 

By performing a sequence of local basis transformations and projection 
onto one of the two fusion channels for each pair of neighboring anyons, 
the resulting two-anyon interaction Hamiltonian can be written as a sum
of local 3-site operators $H=J_2\sum_i H^i_{\rm 2}$ which take the
explicit form
(`$i$' denotes the first in a triple of adjacent sites)
\begin{eqnarray}
H^i_{\rm 2}  & = & - \proj{1\tau1} - \phi^{-2} \proj{\tau1\tau} - \phi^{-1} \proj{\tau\tau\tau}
\\ \nonumber 
                      & & -\phi^{-3/2} \left( \ket{\tau1\tau}\bra{\tau\tau\tau}+{\rm h.c.}
                     \right) ~,
\label{Eq:GoldenChain}
\end{eqnarray}
where $\proj{a}$ projects onto the state $\ket{a}$, e.g.
$\proj{1\tau1} = \ket{1\tau1}\bra{1\tau1}$ and
$\phi=(1+\sqrt{5})/2$ is the golden ratio \cite{GoldenChain}.

Here we want to explore a larger space of models than that given
by this uniform chain with only nearest-neighbor two-anyon interactions. 
One way is to let the strength $J_2$ of the interaction alternate along 
the chain, as illustrated in Fig.~\ref{Fig:Models}a). 
Two chains can be coupled to form a two-leg ladder. 
Another way is to add a spatially-uniform three-anyon interaction, 
as indicated in Fig.~\ref{Fig:Models}b), which because of its rich
phase diagram we discuss in the following.


\paragraph{Majumdar-Ghosh (MG) chain.--}

\begin{figure}[t]
  \includegraphics[width=\columnwidth]{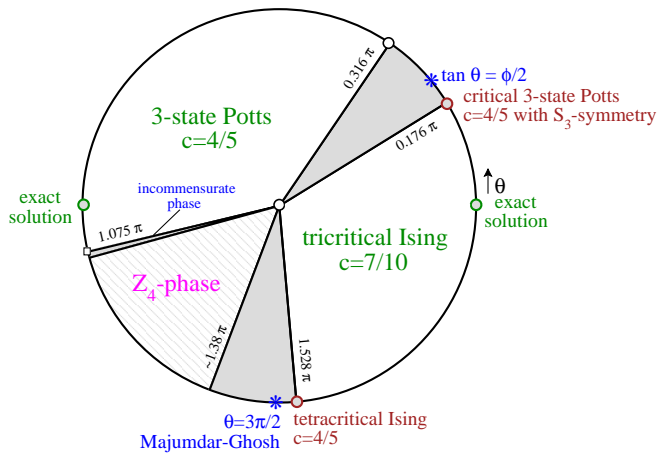}
  \caption{
    (color online)
    The phase diagram of our anyonic Majumdar-Ghosh chain on the circle parameterized by $\theta$,      
    with pairwise fusion term $J_2=\cos\theta$
    and three-particle fusion term $J_3=\sin\theta$.
    Besides extended critical phases around the exactly solvable points ($\theta=0,\pi$) that can
    be mapped to the tricritical Ising model and the 3-state Potts model, there are two gapped
    phases (grey filled).
    The phase transitions (red circles) out of the tricritical Ising phase
    exhibit higher symmetries and are both described by CFTs with central charge $4/5$.
    In the gapped phases exact ground states are known at the positions marked by the stars.
    In the lower left quadrant a small sliver of an incommensurate phase occurs and a phase
    which has $Z_4$-symmetry.  These latter two phases also appear to be
    critical.
   }
  \label{Fig:J3-Chain:PhaseDiagram}
\end{figure}

Three $SU(2)$ spin-1/2's can combine to a total spin 3/2 or 1/2.
For a uniform $SU(2)$ spin-1/2 chain, Majumdar and Ghosh showed 
that an AFM coupling (favoring total spin 1/2) for each set of three
neighboring spins gives rise to a gapped phase with the two
possible dimer coverings being the exact ground states
\cite{MajumdarGhosh}. 
In the same spirit we have asked what possible
phases can be stabilized by a spatially uniform three-particle
interaction term  in our `anyonic generalization of the $SU(2)$ 
Heisenberg model', i.e. a term that
energetically favors each set of three adjacent Fibonacci anyons
to fuse together \cite{MGfootnote}
into either a $1$ or a $\tau$, as illustrated in Fig.~\ref{Fig:Models}b). 
Like the pairwise interaction term such a
three-particle interaction term respects both the translational and
topological symmetries.  We find that the energetic competition
between such two- and three-anyon interactions gives rise to the
rich ground-state phase diagram shown in
Fig.~\ref{Fig:J3-Chain:PhaseDiagram}, which we discuss in some
detail in the following. Similar to the derivation for the pairwise 
interaction term (\ref{Eq:GoldenChain}) we can obtain a local
form $H = J_3\sum_i H^i_{\rm 3}$ of the three-anyon 
interaction term by
a sequence of basis transformations and projections which then
takes the explicit form of a 4-site interaction between
consecutive labels along the chain of fusions
\begin{align}
H^i_{\rm 3} &= \proj{\tau1\tau1} + \proj{1\tau1\tau} +
\proj{\tau\tau\tau1} + \proj{1\tau\tau\tau} +
                           2 \phi^{-2} \proj{\tau\tau\tau\tau} + \nonumber \\
                     & \phi^{-1} \left( \proj{\tau1\tau\tau} + \proj{\tau\tau1\tau} \right)
                           -\phi^{-2} \bigl(\ket{\tau\tau1\tau}\bra{\tau1\tau\tau}+{\rm h.c.}\bigr) + \nonumber \\
                     & \phi^{-5/2} \left( \ket{\tau1\tau\tau}\bra{\tau\tau\tau\tau}
                        +\ket{\tau\tau1\tau}\bra{\tau\tau\tau\tau} {\rm + h.c.} \right) ~,
\label{Eq:J3Chain}
\end{align}
where the site $i$ denotes the first position in each `quad' of
sites. The full Hamiltonian with competing fusion terms then
becomes $H_{J_2,J_3} = \sum_{i} (J_2 H^i_{\rm 2} + J_3 H^i_{\rm
3})$, where we parameterize the couplings by the angle $\theta$ as
$J_2=\cos\theta$ and $J_3=\sin\theta$.  We study periodic chains of
$L$ anyons.

The phase diagram of this model, shown in Fig.~\ref{Fig:J3-Chain:PhaseDiagram}, 
exhibits two critical phases that contain the two exactly solvable points ($\theta=0,\pi$).
These extended critical phases can be described by 2D conformal
field theories and are thereby related to 2D classical
critical points to which an exact mapping was established
at the two solvable points \cite{GoldenChain}. For AFM pair 
interaction ($J_2>0$) this is the tricritical Ising model ($c=7/10$), while
for FM pair interaction ($J_2<0$) it is the critical point of the
3-state Potts model ($c=4/5$). In particular, we note that 
the critical phases found at the exactly solvable points are
{\em stable} upon introducing a small three-anyon fusion term.
While the $J_3$-term respects both translational and topological
symmetries, all translational invariant operators with scaling
dimension $<2$ at the exactly solvable points are found to break
the topological symmetry \cite{GoldenChain}. This shows that the
topological symmetry {\em protects} the gaplessness in the vicinity  
of these points, somewhat analogously to the much-discussed notion
that a topological symmetry protects a ground-state degeneracy in
a gapped topological phase in 2D space.

\begin{figure}[t]
  \includegraphics[width=\columnwidth]{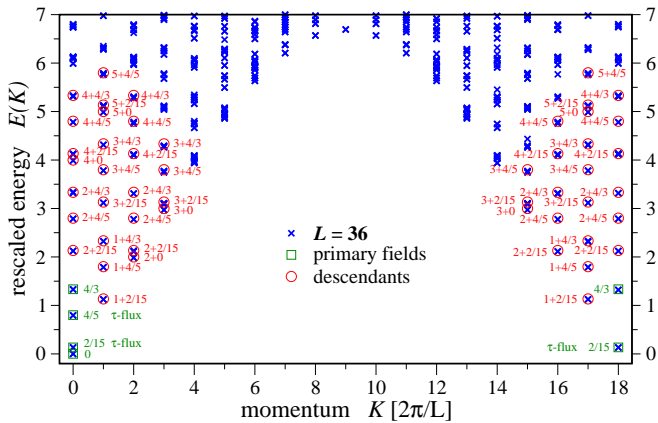}
  \caption{
    (color online)
    Energy spectrum at the $S_3$-symmetric point ($\theta \cong 0.176\pi$) of the
    Majumdar-Ghosh chain.
    The energies have been shifted and  rescaled so that the two lowest eigenvalues
    match the CFT scaling dimensions.
    The open boxes indicate the positions of the primary fields of the parafermionic subset
    of the $c=4/5$ CFT; fields with topological flux $\tau$ are marked.
    The open circles give the positions of multiple descendant fields as indicated.
   }
  \label{Fig:J3-Chain:Theta0176}
\end{figure}

For large three-particle interactions these critical phases
eventually give way to other phases, such as the two distinct
gapped phases indicated by the grey shaded arcs in the phase
diagram. Remarkably, the transition to the gapped phase from the
tricritical Ising phase when both interaction terms are AFM ($J_2>0$
and $J_3>0$) apparently has an ``emergent'' $S_3$ (3-state Potts)
symmetry. Our numerical analysis shows that this transition occurs
at $\theta\cong 0.176\pi$ and is described by the parafermion 
CFT with central charge $c=4/5$, indicative of an additional
$S_3$-symmetry at this point. Fig.~\ref{Fig:J3-Chain:Theta0176}
shows the rescaled energy spectrum at this critical point 
whose (universal) low-energy part is in spectacular agreement with the 
CFT predictions. 
Note the relevant operator with zero momentum, zero flux and scaling
dimension $4/3$, which breaks the $S_3$-symmetry.
It is the leading operator present in the Hamiltonian away from this 
special point, and drives the system into either the gapped or the 
tricritical Ising phase.
In the gapped phase, the topological symmetry is spontaneously 
broken and the resulting ground state, which has
zero total momentum, is two-fold degenerate in the thermodynamic
limit.  
In the tricritical Ising phase, the $Z_2$ sublattice-symmetry breaking 
order parameter corresponds to a more relevant continuum operator 
than the topological order parameter,  while it is the state corresponding 
to the $Z_2$ order parameter which acquires a higher energy in the 
gapped phase, where only the topological symmetry is broken.
At the transition, both order parameters are degenerate, see
Fig.~\ref{Fig:J3-Chain:Theta0176}, and together they form the
order parameter of a critical 3-state Potts model with
$S_3$-symmetry.

\begin{figure}[t]
  \includegraphics[width=\columnwidth]{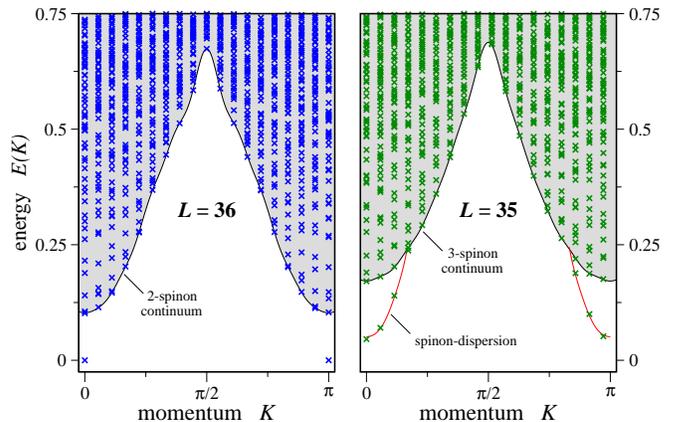}
  \caption{
    (color online)
    Energy spectra at the generalized Majumdar-Ghosh point with ferromagnetic
    three-anyon interaction term only ($\theta=3\pi/2$).
    With both translational and topological symmetry spontaneously broken there
    are four ground states for even length chains (left panel) with two states each
    at momenta $K=0$ and $K=\pi$.
    The gapped low-energy spectrum is a continuum of two-spinon scattering states
    (grey shaded).
    For odd system size (right panel) a distinct single-spinon dispersion
    emerges below a gapped continuum of 3-spinon scattering states.
  }
  \label{Fig:J3-Chain:MGSpectrum}
\end{figure}

In the case of FM three-particle interaction $J_3<0$ the transition at
$\theta \cong -0.472\pi$ between tricritical Ising and gapped
phases is described in terms of CFT by the full $c=4/5$ minimal
model representing \cite{Cardy1986} the {\it tetra}critical Ising model. 
Again we have unambiguously identified the CFT description of this critical
endpoint by assigning the low-energy states in the energy spectrum
similar to Fig.~\ref{Fig:J3-Chain:Theta0176} (not depicted here).
In particular, the topological symmetry forces the system onto the
integrable renormalization group trajectory \cite{RGFlow} into the 
gapped phase or tricritical Ising fixed point, driven again by the relevant 
operator with dimension $4/3$. 
At this tetracritical Ising transition into the gapped phase
the system spontaneously breaks {\it both} the translational and
topological symmetries. As a consequence, we observe a four-fold
ground-state degeneracy throughout this gapped phase for chains
with even length. The nature of this gapped phase is best
characterized at the point $\theta=3\pi/2$ ($J_3=-1$, $J_2=0$)
that constitutes the anyonic analog of the Majumdar-Ghoshpoint
of the spin-$1/2$ Heisenberg chain. At this point the four ground
states for even $L$ take the exact form
\begin{eqnarray}
\label{dimnoflux}
   \ket{\psi_{\rm no-flux}} & = &
         \ket{\tau_x\tau\tau_x\tau\tau_x\tau \ldots} + \phi^{-1} \ket{\tau1\tau1\tau1 \ldots} \pm
         \nonumber \\ &&
         \ket{\tau\tau_x\tau\tau_x\tau\tau_x \ldots} + \phi^{-1} \ket{1\tau1\tau1\tau \ldots} \\
   \ket{\psi_{\tau\rm-flux}} & = &
         \phi^{-1} \ket{\tau_x\tau\tau_x\tau\tau_x\tau \ldots} - \ket{\tau1\tau1\tau1 \ldots} \pm
         \nonumber \\ &&
         \phi^{-1} \ket{\tau\tau_x\tau\tau_x\tau\tau_x \ldots} - \ket{1\tau1\tau1\tau
         \ldots} ~,
   \label{Eq:GSMajumdarGhosh}
\end{eqnarray}
where $\tau_x = \phi^{-1} \ket{1} + \phi^{-1/2} \ket{\tau}$
denotes a normalized superposition of the states $\ket{1}$ and
$\ket{\tau}$ on a single site.  
Note these ground states have total momenta $K=0$ or $K=\pi$,
indicating the two-sublattice ordering.  There are two states at
each momentum, one with a $\tau$-flux and the other without.  Of
course, we can instead make the simpler linear combinations of
these ground states that explicitly break both the
topological and sublattice symmetries: 
these four states are $\ket{\tau_x\tau\tau_x\tau\tau_x\tau
\ldots}$, $\ket{\tau1\tau1\tau1 \ldots}$ and the equivalent states
under translation by one site. 
Note that the density of $1$'s 
[which for these states is $1/(2\phi^2)$ and $1/2$, respectively]
is a simple order parameter that reflects the topological symmetry
breaking.
A pair of these latter states also becomes the exact ground states
if the Hamiltonian explicitly breaks the sublattice symmetry by
alternating the two-anyon interaction along the chain as shown in
Fig.~\ref{Fig:Models}a). For a uniform chain in the tricritical
Ising phase this alternation corresponds to a relevant operator 
that immediately drives the system into a gapped phase. 
On the other hand, the uniform chain in the Potts critical phase has 
$Z_3$ sublattice symmetry and no relevant operators with 
spatial period two, so this chain remains critical when a small alternation 
is added \cite{long}.

The low-energy excitations in the gapped phase around the MG point
are domain walls between the two sublattice-ordered ground states
with a low density of $1$'s -- similar to spinon states in the
spin-$1/2$ MG chain.  For chains of odd $L$ the periodic boundary
conditions force the presence of an odd number of such
spinons/walls, and indeed we observe a distinct single-spinon
dispersion relation over part of the Brillouin zone, 
as shown in the right panel of
Fig.~\ref{Fig:J3-Chain:MGSpectrum}. Explicitly evaluating the
topological symmetry operator \cite{GoldenChain}
we find that these one-spinon states all have
a $\tau$-flux. For even length rings we observe a continuum of
two-spinon scattering states in the spectrum shown in the left
panel of Fig.~\ref{Fig:J3-Chain:MGSpectrum}.

\begin{figure}[t]
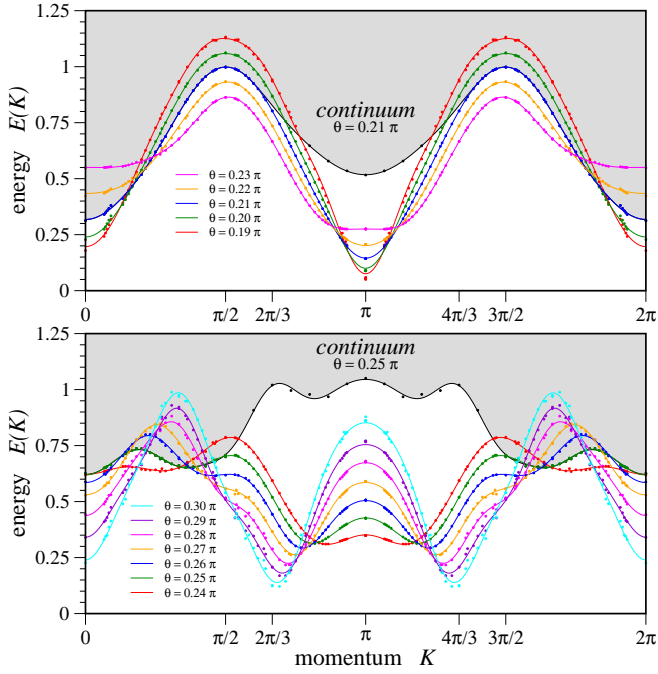

  \includegraphics[width=\columnwidth]{./Spectrum1.eps}
  \includegraphics[width=\columnwidth]{./Spectrum2.eps}
  \caption{
    (color online)
    Schematic energy spectra in the gapped phase for $0.19 \leq \theta/\pi \leq 0.30$.     
    Below a continuum of scattering states (grey shaded) a distinct
    quasiparticle band forms.
    The plot combines data with $24 \leq L \leq 36$.
    The solid lines are a guide to the eye.
   }
  \label{Fig:J3-Chain:GappedPhase}
\end{figure}

At the $S_3$-symmetric transition out of the tricritical Ising
phase ($\theta\cong 0.176\pi$) the system spontaneously breaks
only the topological symmetry resulting in a two-fold degenerate
ground state at total momentum $K=0$ in the gapped phase.
With increasing coupling $J_3$, away from the transition,  
an energy gap opens in the spectrum, with a distinct
quasiparticle dispersion forming below a continuum of scattering
states as shown in Fig.~\ref{Fig:J3-Chain:GappedPhase}. Explicitly
evaluating the topological symmetry operator \cite{GoldenChain}
we find that these quasiparticle states all have a $\tau$-flux. 
For small $J_3$, the
lowest energy quasiparticle remains at momentum $K=\pi$, with both
the gap and the mass increasing with increasing $J_3$. Near
$\theta \cong 0.24\pi$ the gap reaches a maximum value and the
mass diverges. The minimum of the quasiparticle dispersion
bifurcates and continuously moves away from $K=\pi$, see the lower
panel in Fig.~\ref{Fig:J3-Chain:GappedPhase}. Eventually, the two
minima approach the commensurate momenta $K=2\pi/3, 4\pi/3$ and as
these modes soften the system enters the 3-state Potts critical
phase at $\theta \cong 0.316\pi$. Similar to the MG point we can
identify a point in this gapped phase at which we can determine
the exact form of the two ground states. For
$J_3=\frac{\phi}{2}J_2$, we note that the off-diagonal term in
Hamiltonian (\ref{Eq:GoldenChain}) is exactly cancelled by its
counterpart in the $J_3$-Hamiltonian (\ref{Eq:J3Chain}), and at
this angle, $\tan{\theta}=\phi/2$, the ground states take the
explicit form
\begin{eqnarray}
  \ket{\psi_{\rm no-flux}} & = &
    \ket{ \tilde{\tau}_x \tilde{\tau}_x \tilde{\tau}_x \tilde{\tau}_x \ldots}
      +(-1)^L\phi^{-1} \ket{\tau\tau\tau\tau \ldots} \\
  \ket{\psi_{\tau\rm-flux}}       & = & \phi^{-1} \ket{ \tilde{\tau}_x \tilde{\tau}_x \tilde{\tau}_x \tilde{\tau}_x \ldots}
      -(-1)^L \ket{\tau\tau\tau\tau \ldots} \,
\end{eqnarray}
where $\tilde{\tau}_x = \phi^{1/2} \ket{1} +\phi^{-1} \ket{\tau}$
is a single site superposition. 
(A projector onto the states in the Hilbert space is implicitly assumed.)
Again note that the density of $1$'s is a simple order
parameter for topological symmetry breaking.

Finally, we note that when both interaction terms are FM 
the critical 3-state Potts phase gives way to a small sliver of an
incommensurate phase and then a phase with $Z_4$ sublattice
symmetry. All of these phases appear to be critical or nearly critical.
In the incommensurate phase, correlations in the local density of $1$'s
oscillate with a spatial period varying between 3 and 4 lattice
spacings.


\paragraph{Breaking the topological symmetry.--}
To further elucidate the role of the topological symmetry in
determining the nature of the observed phases, we can {\em
explicitly} break this symmetry in a modified Hamiltonian.  Since
the number of 1's, 
$N_1$, is a simple order parameter for this symmetry, we add a
term $hN_1$ to the Hamiltonian, where $h$ is the field that breaks
the symmetry.
Within the gapped phases the ground-state degeneracy is
immediately lifted by $h \neq 0$, with the state(s) with lower
(higher) density
of 1's favored by positive (negative) $h$.  
At the point $\tan{\theta}=\phi/2$ these states are precisely the ``vacuum"
state of 1's $\ket{\tau\tau\tau\tau \ldots}$ ($N_1=0$) for $h>0$ and the
state $\ket{ \tilde{\tau}_x \tilde{\tau}_x \tilde{\tau}_x
\tilde{\tau}_x \ldots}$ ($N_1>0$) for $h<0$. The whole gapped phase for AFM
$J_2$ and $J_3$ can thus be identified as a first-order transition
with a `liquid-gas' coexistence of 1's at $h=0$.  
A similar picture, but with broken sublattice symmetry, holds
for the gapped phase around the MG point. Varying 
$h$ in the {\it critical} phases of our phase diagram, there is a
weaker feature in the ground-state energy $E(h)$: the
`topological susceptibility' $\chi_t(h) \propto d^2E/dh^2$
diverges with system size at $h=0$. This indication of a
continuous phase transition on varying $h$ confirms that the
topological symmetry is a full participant in the critical
behavior at $h=0$.

In conclusion, the anyonic quantum chains possess a topological symmetry
that forces their corresponding 2D classical models onto a highly fine-tuned 
submanifold of their respective phase diagrams.
The competition of anyonic exchange interactions allows one to move within this 
manifold containing a plethora of both, (multi)critical phases such as tricritical Ising, 
3-state Potts, tetracritical Ising, and various gapped phases with  spontaneously 
broken topological symmetry including an analog of the Majumdar-Ghosh point.

We acknowledge insightful discussions with N. Bonesteel, M. Freedman, A. Kitaev,
M. Levin, and Z. Wang. Our numerical simulations were based on the
ALPS libraries \cite{ALPS}.  
D. A. H. thanks NSF DMR-0213706 and PHY-0551164 for support.
A. W. W. L. was supported, in part, by NSF DMR-0706140.



\begin{thebibliography}{99}

\bibitem{Kitaev03}
A. Kitaev, Ann. Phys. {\bf 303}, 2 (2003).

\bibitem{DasSarma07}
S. Das Sarma, {\it et al.}, arXiv/0707.1889.

\bibitem{Marcus07}
J. B. Miller, {\it et al.}, Nature Physics {\bf 3}, 561 (2007).

\bibitem{MooreRead}
G. Moore and N. Read, Nucl. Phys. B {\bf 360}, 362 (1991).

\bibitem{ReadRezayi}
N. Read and E. Rezayi, \prb {\bf 59}, 8084 (1999).

\bibitem{GoldenChain}
A. Feiguin, {\em et al.},  Phys. Rev. Lett. {\bf 98}, 160409
(2007).

\bibitem{Xia04}
J. S. Xia, {\em et al.}, Phys. Rev. Lett. {\bf 93}, 176809 (2004).

\bibitem{MGfootnote}
Like for $SU(2)$ spin-1/2 chains the 3-anyon interaction is
a sum of  nearest and next-nearest neighbor pairwise interactions
\cite{long}.

\bibitem{long}
S. Trebst, {\em et al.} (in preparation).

\bibitem{Huse}
D. A. Huse, Phys. Rev. Lett. {\bf 49}, 1121 (1982); Phys. Rev. B
{\bf 30}, 3908 (1984).

\bibitem{MajumdarGhosh}
C. K. Majumdar and D. K. Ghosh, J. Math. Phys. {\bf 10}, 1399
(1969).

\bibitem{Cardy1986}
J. L. Cardy, Nucl. Phys. {\bf B270}, 186 (1986).

\bibitem{RGFlow} 
A. W. W. Ludwig, J. L. Cardy, Nucl. Phys. B {\bf 285}, 687 (1987);
A. B. Zamolodchikov, Sov. J. Nucl. Phys. {\bf 46}, 1090 (1987);
A. B. Zamolodchikov in {\it
Integrable systems in quantum field theory and statistical mechanics};
eds. M. Jimbo, T. Miwa, A. Tsuchiya (Avademic Press, Tokyo, 1989).

\bibitem{ALPS}
A.F. Albuquerque, {\em et al.}, J. of Magn. and Magn. Materials
{\bf 310}, 1187 (2007).

\end{thebibliography}
\end{document}